\documentstyle[twoside,fleqn,espcrc2,epsf]{article}

   \newcommand{\cqq}{ {\cal Q}^{\dagger}{\cal Q} }

\newcommand{\AmS}{{\protect\the\textfont2
  A\kern-.1667em\lower.5ex\hbox{M}\kern-.125emS}}

\title{Hubbard Model with L\"{u}scher fermions}

\author{P. Sawicki\address{Institute of Physics,
        Jagellonian University, \\
        Reymonta 4, Cracow, Poland}%
        \thanks{Supported by the KBN grant no PB 0488/93/P3.}
        and
        J. Wosiek\address{Institute of Computer Science, Jagellonian University
, \\
        Reymonta 4, Cracow, Poland}
        \thanks{Presented by J. Wosiek.} }

\begin{document}

\begin{abstract}
First applications of the new algorithm simulating dynamical fermions are
reported. The method reproduces previous results obtained with
different techniques.
\end{abstract}

\maketitle

\section{DYNAMICAL FERMIONS AND L\"{U}SCHER REPRESENTATION}
The problem of dynamical fermions, even though solved in principle,
still provides the considerable challenge in practice. Existing
exact algorithms require huge computing resources due to the
strong correlations between generated configurations \cite{ken}. Therefore
the recent proposal of L\"{u}scher has attracted a lot of interest
 \cite{l1},\cite{l2}. To remind, the main difficulty with simulating the
theory with dynamical fermions consists of the nonlocality
of the fermionic determinant
\begin{equation}
\int [d\Psi d\overline{\Psi}]  \exp{(-\overline{\Psi} {\cal M} \Psi) }
\sim  det({\cal M}),
\end{equation}
which depends functionally on the gauge field in the case of QCD
for example.
 The local and positive representation for the positive powers
of $det({\cal M})$ was not found up to date. L\"{u}scher method
exploits the well known fact that the {\em inverse} of the determinant
{\em has} such a representetions in terms of the standard bosonic
fields. He therefore proposes to use the polynomial approximation
of the reciprocal function
\begin{equation}
 {1\over s}=\lim_{N\rightarrow\infty} P_N(s).  \label{poly}
\end{equation}
In the simplest case $P_N(s)$ can be taken as the geometrical series
(in $1-s$),
in practice better choices are known. Decomposing $P_N$ into the product
form and using the symmetries of roots of real polynomials
 L\"{u}scher proves that the partition function of the full QCD with
two flavours of dynamical quarks has positive and local representation
in terms of the original gauge fields and $N$ bosonic fields associated
with zeros of $P_N(s)$.

It remains now to be tested if this elegant trick is also advantageous
in practice. In particular the crucial question is: how do the parameters of
this representation scale with the volume and with the bare parameters
of the original theory in its critical region?  All these important
and critical problems are being now thouroghly checked \cite{jag}.

In this contribution I report on the first results of applying the
L\"{u}scher's method to the (Euclidean) three-dimensional Hubbard model,
which is the convenient and nontrivial testing ground of the new techniques
dealing with dynamical fermions.

\section{THE HUBBARD MODEL AND ITS BOSONIC REPRESENTATION }

The Hubbard model is a many-body theory of nonrelativistic electrons
with the nonlinear point-like coupling representing the effective
Coulomb interactions \cite{frad},\cite{mo}. Its particular Euclidean
formulation contains
also the continuous Hubbard-Stratonovich field which couples to
the electrons, hence may be regarded as analogous to the Yang-Mills
fields of QCD. Even though the nonrelativistic, the model contains all
the numerical difficulties caused by the fermionic statistcs, i.e. by the
Pauli exclusion principle. Being simpler than QCD it still describes
interesting
physics of a nontrivial many-body system.

Hubbard Hamiltonian may be written as
\begin{eqnarray}
\lefteqn{  H=-K\sum_{<ij>\sigma} a_{i\sigma}^{\dagger} a_{j\sigma}
-{U\over 2}\sum_i (n_{i\uparrow} - n_{j\downarrow})^2  } \\
  & & + \mu \sum_{i\sigma} n_{i\sigma},
\end{eqnarray}
where $n_{i\sigma}=a^{\dagger}_{i\sigma} a_{i\sigma}$, and $a_{i\sigma}$
denotes the creation operator of an electron at the lattice site $i$
and with the spin $\sigma=\uparrow,\downarrow$.
$K, U$ and $\mu$  are  the hopping parameter, coupling constant and
the chemical potential respectively. Standard transfer matrix formalism
gives the following Euclidean representation for the partition function
\cite{cr}
\begin{equation}
Z=\int [dA] \exp{(-\sum_{i,t} A_{it}^2/2)}
\det{{\cal M}_{+}} \det{{\cal M}_{-}},
\end{equation}
where the continuous Hubbard-Stratonovich field $A_{it}$ was introduced
to decouple the quartic interaction term. The discrete index
$t=1,\dots , N_t$ labels the time slices.
Fermionic matrices ${\cal M}_{\pm}$ define the bilinear fermionic actions
\begin{eqnarray}
\Psi^{\ast} {\cal M}_{\pm} \Psi = {K\beta\over N_t}
\sum_{<ij>t} \Psi^{\ast}_{it}\Psi_{jt} +
\sum_{it}\Psi^{\ast}_{it}(\Psi_{it}-\Psi_{it-1})   \nonumber \\ +
\sum_{it}\Psi^{\ast}_{it}\Psi_{it}
\left(
       \exp{
            \left[
\sqrt{U\beta\over N_t} A_{it} - (U\pm\mu){\beta\over N_t}
           \right]
            }
-1
\right). \nonumber
\end{eqnarray}
At $\mu=0$, which corresponds to the half-filling,
\begin{equation}
\det{\cal M_{+}} \det{\cal M_{-}} = \det{\cal M}^2 =
\det{{\cal M}{\cal M}^{\dagger}},
\end{equation}
because the determinants are real. We can therefore apply the L\"{u}scher
trick and write the polynomial approximation for the inverse
\begin{eqnarray}
\lefteqn{ {1\over\cqq}= P_{2N}(\cqq) = \prod_{k=1}^{2N} (\cqq-z_k) }
   \label{prod} \\
 & &   =\prod_{k=1}^{N} (\cqq-\alpha_k-i\beta_k)
                             (\cqq-\alpha_k+i\beta_k) ,  \nonumber
\end{eqnarray}
 since the roots $z_k\equiv \alpha_k+i\beta_k$ of the polynomial
 $P_{2N}(z)$ exist in the complex conjugate pairs.
${\cal Q^{\dagger} \cal Q}\equiv
 {\cal M ^{\dagger} \cal M }/\lambda_{max}$ with $\lambda_{max}$ being the
 largest eigenvalue
 of ${\cal M^{\dagger} \cal M}$.

We therefore have the local and positive representation for the
partition function of the Hubbard model.
\begin{eqnarray}
  \lefteqn{   Z  \simeq  \int   [dA d\phi]
                                            \exp{ \left( -\int    A^2(x)/2 \;
     d^3x \right)  }   } \\
   \lefteqn{   \exp{  \left(         -\int  \sum_{k=1}^{N}
\phi^{\dagger}_k
                \left[
                       (\cqq-\alpha_k)^2+\beta_{k}^{2}
               \right]
    \phi_k               d^3 x  \right)
                       }    }.  \nonumber  \\
                       \label{zet}
\end{eqnarray}
where we have used continuous notation for simplicity.
Hence the original system of nonlinearly interacting fermions
was replaced by $N$ complex bosonic fields coupled to the single
real scalar field $A(x)$. This can be simulated with the standard local
Monte Carlo techniques. Compared to the original L\"{u}scher
mapping we have half as many fields $\phi_k(x)$, however our bosonic
action is more complicated since $\cqq$  contains also
next-to-nearest-neighbour interactions.

\subsection{IMPLEMENTATION AND RESULTS}

Convergence of the polynomial approximation, Eq.(\ref{poly}), cannot be
uniform in $s$ in the whole interval (0,1). Hence L\"{u}scher introduces
the cut-off $\epsilon > 0$ such that the series (\ref{poly}) is uniformly
convergent for $ \epsilon < s < 0 $. This parameter enters directly into
the construction of the optimal polynomials $P_N$. The value of $\epsilon$
is crucial for the practical applicability of the algorithm since it controls
the number of bosonic fields required to reach the prescribed accuracy.
Ideally $\epsilon$ should be smaller that the smallest eigenvalue of
the normalized matrix $\cqq$. Otherwise corrections for a few
eigenvalues lower than $\epsilon$ could be implemented \cite{l2}.

\begin{figure}[htb]
\vspace{9pt}
\epsfxsize=6.5cm \epsfbox{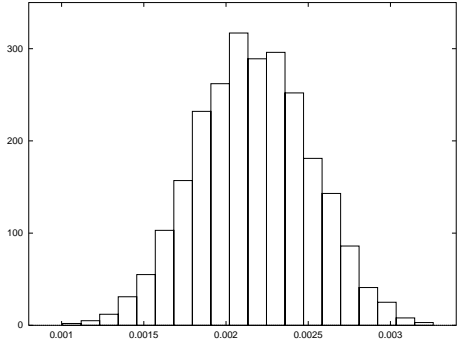}
\caption{Distribution of the smallest eigenvalue of the matrix $\cqq$
    found in the simulation described in text. }
\label{fig:largenenough}
\end{figure}

\begin{figure}[htb]
\vspace{9pt}
\epsfxsize=6.5cm \epsfbox{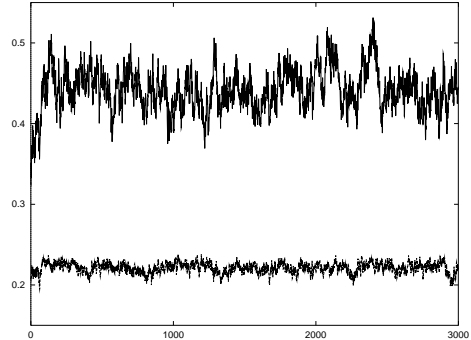}
\caption{The history of run of total length 3000 events. The larger
         values are for density of electrons, the smaller for density
         of pairs.}
\label{fig:toosmall}
\end{figure}

Our preliminary simulations were done on the $5^3$ lattice at $K=1$
and $U=1$. These parameters were chosen to allow comparison with the
earlier Creutz results obtained with different algorithm \cite{cr}.
One Monte Carlo sweep consisted of the heatbath generation of all fields
$\phi_k$ and the Metropolis update of the Hubbard-Stratonovich field $A(x)$.
Fig. 1 shows the distribution of the minimal eigenvalue of $\cqq$.
On this basis we have chosen $\epsilon=0.001$. The number of auxiliary
fields $\phi_k$ was fixed to $N=100$. This guaranteed that the relative
error introduced by finite $N$ in Eq.(\ref{poly}) was smaller than $10^{-4}$
for $\epsilon < s$. Instead of $\lambda_{max}$ needed to define the
normalized matrix ${\cal Q}$ we have used the upper bound
$\overline{\lambda}$ which can be readily derived
\begin{equation}
\lambda_{max} < 2 [\lambda_{max}(A^{\dagger} A) +
               max(D)^2 ] \equiv\overline{\lambda},
\end{equation}
where $D$ is the diagonal part of ${\cal M}=A+D$. Hubbard-Stratonovich field
was cut-off by $A(x) < 4$ which would correspond to the four standard
deviations if the determinants were neglected. This bound was never
violated in practice.
Fig.2 shows the history of the density of the electrons $n_{\uparrow}$
and of the local number of pairs $n_{\uparrow}(x)n_{\downarrow}(x)$.
Both observables stabilize relatively quickly and the equilibrium averages
agree with the results quoted by Creutz \cite{cr} for slightly bigger
lattice. For the $6^2\times 8$ lattice we obtain $<n_{\uparrow}>=0.47(3)$
and $<n_{\uparrow} n_{\downarrow}>=0.220(3)$ while the corresponding
numbers read from the Creutz plots are $0.48$ and $0.22$ with the
read-out error $\sim 0.01$.
 The run summarized in Fig.2 took about 10 hrs of the HP-720.

We have found however that the configurations generated by this algorithm
are strongly correlated. The autocorrelation time measured on both densities
is approximately 100 sweeps. Similar phenomenon was also reported at this
Conference by Jegerlehner while testing QCD implementation \cite{jag}.

In conclusion, the Hubbard Model is in the L\"{u}scher class, i.e. it
can be mapped onto a system of bosons with local interactions.  The half-
filed case has a positive Boltzman factor which admits standard Monte
Carlo approach. Further studies are needed to investigate whether
the large autocorrelation times can be eliminated.

We thank Andr\'e\  Morel for the numerous and instructive discussions.
This work was supported by the KBN grant no PB 0488/93/P3.

\end{document}